\documentclass[a4paper,12pt]{article}
\usepackage{latexsym}
\usepackage[cp866]{inputenc}
\usepackage[english]{babel}

\input epsf

\begin{document}

\begin{center}
\textbf{TOWARDS STUDY OF LIGHT SCALAR MESONS IN POLARIZATION
PHENOMENA}

\vspace{8mm}

N.N.~Achasov$^{\,1\,\dag}$ and G.N.~Shestakov$^{\,1}$

\vspace{4mm}

\begin{small}
(1) \emph{Laboratory of Theoretical Physics, Sobolev Institute for
Mathematics, Academician Koptiug Prospekt, 4, Novosibirsk, 630090,
Russia;} $\dag$ \emph{E-mail: achasov@math.nsc.ru}
\end{small}
\end{center}

\vspace{0.0mm} 

\begin{abstract}
After a short review of the production mechanisms of the light
scalars which reveal their nature and indicate their quark
structure, we suggest to study the mixing of the isovector
$a^0_0(980)$ with the isoscalar $f_0(980)$ in spin effects.
\end{abstract}

\vspace{2mm}

\noindent\textbf{1\ \ Introduction.}

The scalar channels in the region up to 1 GeV became a stumbling
block of QCD. The point is that both perturbation theory and sum
rules do not work in these channels because there are not solitary
resonances in this region.

As  Experiment suggests, in chiral limit confinement forms
colourless observable hadronic fields and spontaneous breaking of
chiral symmetry with massless pseudoscalar fields. There are two
possible scenarios for QCD realization at low energy: 1.
$U_L(3)\times U_R(3)$ non-linear $\sigma$ model, 2. $U_L(3)\times
U_R(3)$ linear $\sigma$ model. The experimental nonet of the light
scalar mesons suggests $U_L(3)\times U_R(3)$ linear $\sigma$ model.

\vspace{2mm}

\noindent\textbf{2\ \ \boldmath $SU_L(2)\times SU_R(2)$ Linear
$\sigma$ Model, Chiral Shielding in $\pi\pi\to\pi\pi$ \cite{AS9407}}

Hunting the light  $\sigma$ and $\kappa$ mesons had begun in the
sixties. But the fact that both $\pi\pi$ and $\pi K$ scattering
phase shifts do not pass over $90^0$ at putative resonance masses
prevented to prove their existence in a conclusive way.

Situation changes when we showed that in the linear $\sigma$ model
there is a negative background phase which hides the $\sigma$ meson
\cite{AS9407}. It has been made clear that shielding wide lightest
scalar mesons in chiral dynamics is very natural. This idea was
picked up and triggered new wave of theoretical and experimental
searches for the $\sigma$ and $\kappa$ mesons.

Our approximation is as follows (see Fig.\,1):
$T_0^{0(tree)}$\,=\,$\frac{m_\pi^2-m_\sigma^2}{32\pi f^2_\pi}\left[
5-3\frac{m_\sigma^2-m_\pi^2}{m_\sigma^2-s}-2\frac{m_\sigma^2-
m_\pi^2}{s-4m_\pi^2}\right.\\ \left.\times\ln\left
(1+\frac{s-4m^2_\pi}{m_\sigma^2}\right)\right]$,\ \
$T^0_0$\,=\,$\frac{T_0^{0(tree)}}{1-i\rho_{\pi\pi}
T_0^{0(tree)}}$\,=\,$\frac{e^{(2i\delta_{bg}+\delta_{res})}-1}{2i\rho_{\pi\pi}}
$\,=\,$\frac{e^{2i\delta^0_0}-1}{2i\rho_{\pi\pi}}$ = $T_{bg}+
e^{2i\delta_{bg}}T_{res}$, \ $T_{res}$
$=\frac{\sqrt{s}\Gamma_{res}(s)/\rho_{\pi\pi}}{M^2_{res}-s+
\mbox{Re}\Pi_{res}(M^2_{res})-\Pi_{res}(s)}$
$=\frac{e^{2i\delta_{res}}-1}{2i\rho_{\pi\pi}}$,\ \
$T_{bg}=\frac{e^{2i\delta_{bg}}-1}{2i\rho_{\pi\pi}}$
$=\frac{\lambda(s)}{1-i\rho_{\pi\pi}\lambda(s)}$,\ \ $\lambda(s)
=\frac{m_\pi^2-m_\sigma^2}{32\pi f^2_\pi}\left[5-2\right.\\
\left.\times\frac{m_\sigma^2-m_\pi^2}{s-4m_\pi^2}\ln\left
(1+\frac{s-4m^2_\pi}{m_\sigma^2}\right)\right]$,\ \
$\mbox{Re}\Pi_{res}(s)$=$-\frac{g_{res}^2(s)}{16\pi}\lambda(s)
\rho_{\pi\pi}^2$,\ \
$\mbox{Im}\Pi_{res}(s)$=$\sqrt{s}\Gamma_{res}(s)$=$
\frac{g_{res}^2(s)\rho_{\pi\pi}}{16\pi}$,\\
$g_{res}(s)$=$\frac{g_{\sigma\pi\pi}}{|1-i\rho_{\pi\pi}\lambda(s)|}$,
\ \ $M^2_{res}$=$m_\sigma^2-\mbox{Re}\Pi_{res}(M^2_{res})$,\ \
$\rho_{\pi\pi}$=$\sqrt{1-\frac{4m_\pi^2}{s}}$,\ \
$g_{\sigma\pi\pi}$=$\sqrt{\frac{3}{2}}g_{\sigma\pi^+\pi^-}$\\
=$\sqrt{\frac{3}{2}}\frac{m^2_\pi-m^2_\sigma}{f_\pi}$;\ \
$T^{2(tree)}_0=\frac{m_\pi^2-m_\sigma^2}{16\pi f^2_\pi}\left
[1-\frac{m_\sigma^2-m_\pi^2}{s-4m_\pi^2}\ln\left
(1+\frac{s-4m^2_\pi}{m_\sigma^2}\right)\right]$,\ \
$T^2_0$=$\frac{T_0^{2(tree)}}{1-i\rho_{\pi\pi}T_0^{2(tree)}}$=$
\frac{e^{2i\delta_0^2}-1}{2i\rho_{\pi\pi}}$.

The results in our approximation are: $M_{res}$ =$0.43$\,GeV,
$\Gamma_{res}(M^2_{res})$\,=\,$0.67$\,GeV, $m_\sigma$ =0.93\,GeV,
$\Gamma^{renorm}_{res}(M^2_{res})$\,=\,$\frac{\Gamma_{res}(
M^2_{res})} {1+d\mbox{Re}\Pi_{res}(s)/ds|_{s=M^2_{res}}}$\,=\,0.53
GeV, $g_{res}(M^2_{res})/g_{\sigma\pi\pi}$=0.33, $a^0_0$=\\ 0.18\,
$m_\pi^{-1}$, $a^2_0$=$-0.04\, m_\pi^{-1}$, the Adler zeros
$(s_A)^0_0$=$0.45\, m^2_\pi$ and $(s_A)^2_0$=$2.02\, m^2_\pi$. The
chiral shielding of the $\sigma(600)$ meson in
$\pi\pi$\,$\to$\,$\pi\pi$ is illustrated in Fig.\,2 with the help of
the $\pi\pi$ phase shifts $\delta_{res}$, $\delta_{bg}$,
$\delta^0_0$ (a), and with the help of the corresponding cross
sections (b).

\begin{figure} \centerline{\epsfysize=1.5in 
\epsfbox{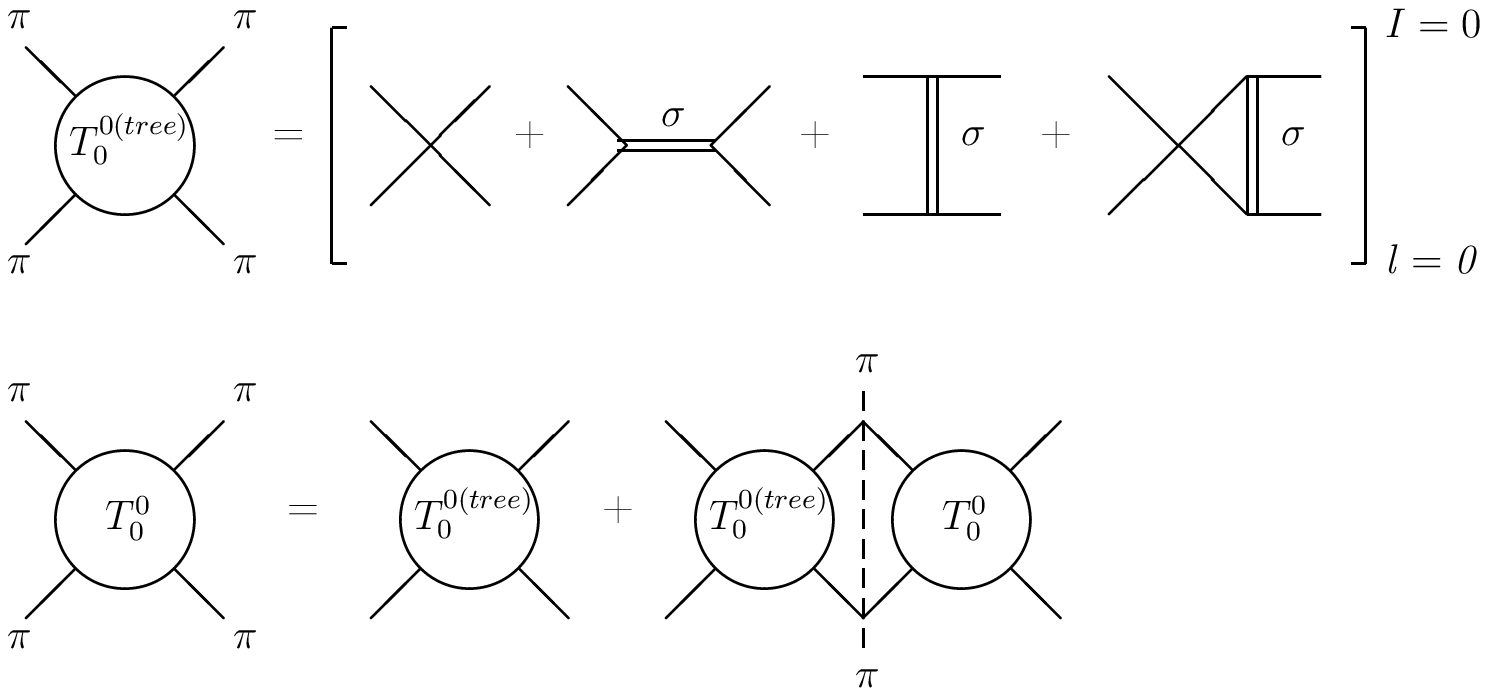}}\vspace{-2mm} \caption {{\footnotesize The
graphical representation of the $S$ wave $I=0$ $\pi\pi$ scattering
amplitude $T^0_0$.}}\end{figure}

\begin{figure}\vspace{3mm} \centerline{\epsfysize=1.6in
\epsfbox{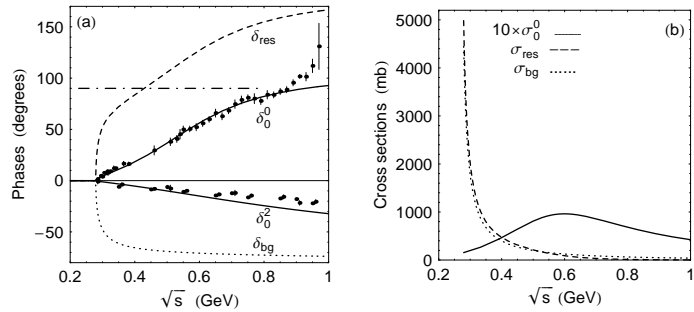}}\vspace{-2mm} \caption {{\footnotesize The
$\sigma$ model. Our approximation.
$\delta^0_0=\delta_{res}+\delta_{bg}$.
($\sigma^0_0,\sigma_{res},\sigma_{bg}$)=$\frac{32\pi}{s}(|T^0_0|^2,
|T_{res}|^2,|T_{bg}|^2)$.}}\end{figure}

\vspace{2mm}

\noindent\textbf{3\ \ The \boldmath{$\sigma$} Propagator
\cite{AS9407}}

$1/D_\sigma(s)$=$1/[M^2_{res}$--$s$+$\mbox{Re}\Pi_{res}(M^2_{res})$--$
\Pi_{res}(s)]$. The $\sigma$ meson self-energy $\Pi_{res}(s)$ is
caused by the intermediate $\pi\pi$ states, that is, by the
four-quark intermediate states. This contribution shifts the
Breit-Wigner (BW) mass greatly $m_\sigma-M_{res}\approx$\,0.50\,GeV.
So, half the BW mass is determined by the four-quark contribution at
least. The imaginary part dominates the propagator modulus in the
region 0.3\,GeV\,$<\sqrt{s}<$\,0.6\,GeV. So, the $\sigma$ field is
described by  its four-quark component at least in this energy
(virtuality) region.

\vspace{2mm}

\noindent\textbf{4\ \ Four-quark Model}

The nontrivial nature of the well-established light scalar
resonances $f_0(980)$ and $a_0(980)$ is no longer denied practically
anybody. As for the nonet as a whole, even a cursory look at PDG
Review gives an idea of the four-quark structure of the light scalar
meson nonet, $\sigma(600)$, $\kappa(700-900)$, $f_0(980)$, and
$a_0(980)$, inverted in comparison with the classical $P$ wave
$q\bar q$ tensor meson nonet $f_2(1270)$, $a_2(1320)$,
$K_2^\ast(1420)$, $\phi_2^\prime(1525)$. Really, while the scalar
nonet cannot be treated as the $P$ wave $q\bar q$ nonet in the naive
quark model, it can be easy understood as the $q^2\bar q^2$ nonet,
where $\sigma$ has no strange quarks, $\kappa$ has the $s$ quark,
$f_0$ and $a_0$ have the $s\bar s$ pair. Similar states were found
by Jaffe in 1977 in the MIT bag \cite{Ja77}.

\vspace{2mm}

\noindent\textbf{5\ \ Radiative Decays of the \boldmath{$\phi$}
Meson and the $K^+K^-$ Loop Model \cite{A8907}}

Ten years later we showed that $\phi$\,$\to$\,$\gamma
a_0$\,$\to$\,$\gamma\pi\eta$ and $\phi$\,$\to$\,$\gamma
f_0$\,$\to$\,$ \gamma\pi\pi$ can shed light on the problem of the
$a_0(980)$ and $f_0(980)$ mesons. Now these decays are studied not
only theoretically but also experimentally. When basing the
experimental investigations, we suggested one-loop model $\phi\to
K^+K^-\to\gamma a_0/f_0$, see Fig. 3. This model is used in the data
treatment and is ratified by experiment, see Fig. 4. Gauge
invariance gives the conclusive arguments in favor of the $K^+K^-$
loop transition as the principal mechanism of the $a_0(980)$ and
$f_0(980)$ meson production in the $\phi$ radiative decays.

\begin{figure}\begin{center}\begin{tabular}{ccc}{\epsfysize=0.7in\epsfbox{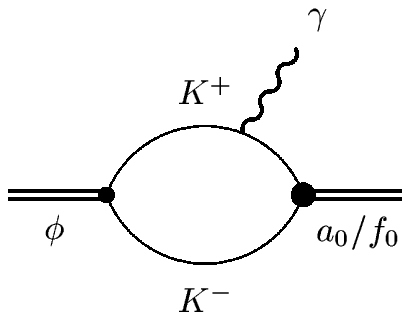}}\
\ & \raisebox{-3.7mm}{\epsfysize=0.7in\epsfbox{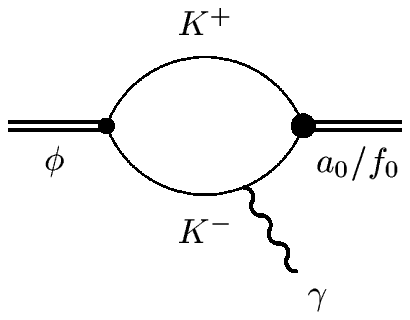}}\ \ &
{\epsfysize=0.7in\epsfbox{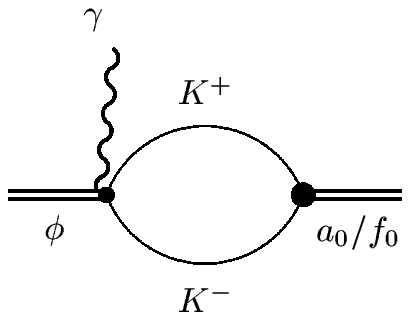}}\end{tabular}\end{center}\vspace{-4mm}\caption{{
\footnotesize The $K^+K^-$ loop model.}}\end{figure}

\vspace{2mm}

\noindent\textbf{6\ \ \boldmath{The $K^+K^-$} Loop Mechanism is
Four-Quark Transition \cite{A8907}}

In truth this means that the $a_0(980)$ and the $f_0(980)$ are seen
in the $\phi$ meson radiative decays owing to the $K^+K^-$
intermediate state. So, the mechanism of the $a_0(980)$ and
$f_0(980)$ production in the $\phi$ meson radiative decays is
established at a physical level of proof. We are dealing with the
four-quark transition. A radiative four-quark transition between two
$q\bar q$ states requires creation and annihilation of an additional
$q\bar q$ pair, i.e., such a transition is forbidden by the OZI
rule, while a radiative four-quark transition between $q\bar q$ and
$q^2\bar q^2$ states requires only creation of an additional $q\bar
q$ pair, i.e., such a transition is allowed by the OZI rule. The
large $N_C$ expansion supports this conclusion.

\begin{figure}\vspace{2mm}\begin{center}\begin{tabular}{cc}{\epsfysize=1.5in
\epsfbox{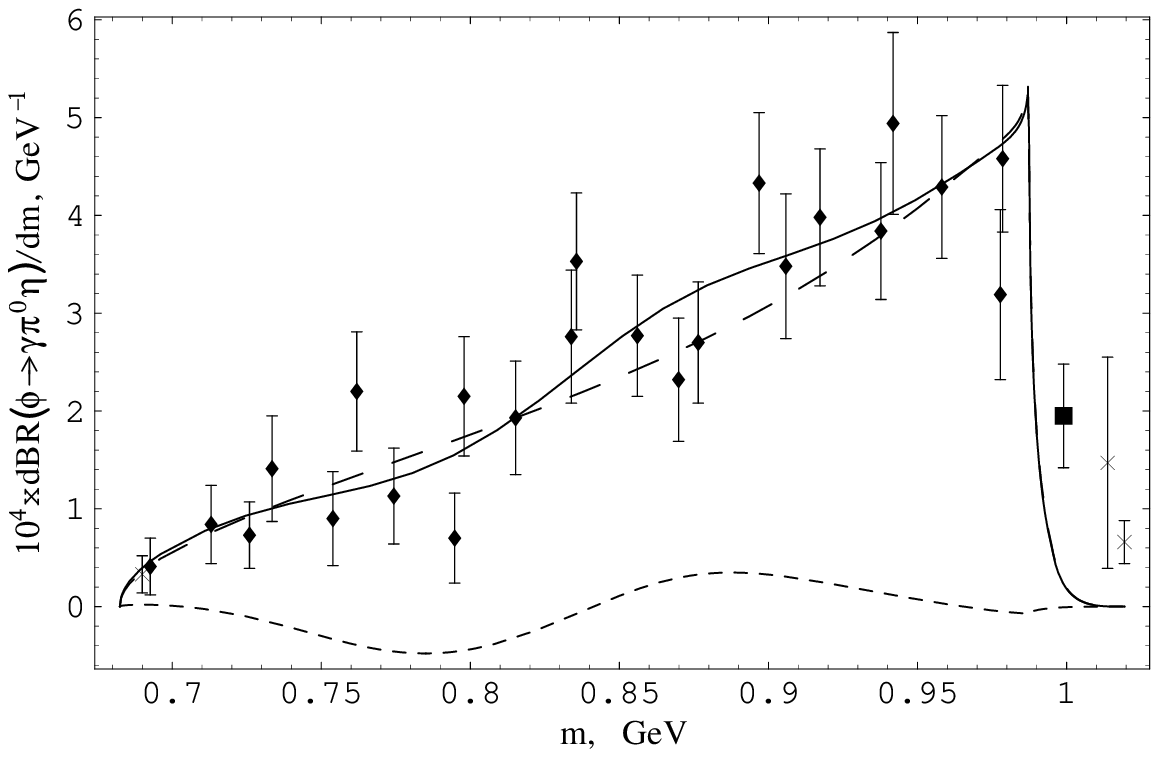}}& {\epsfysize=1.5in\epsfbox{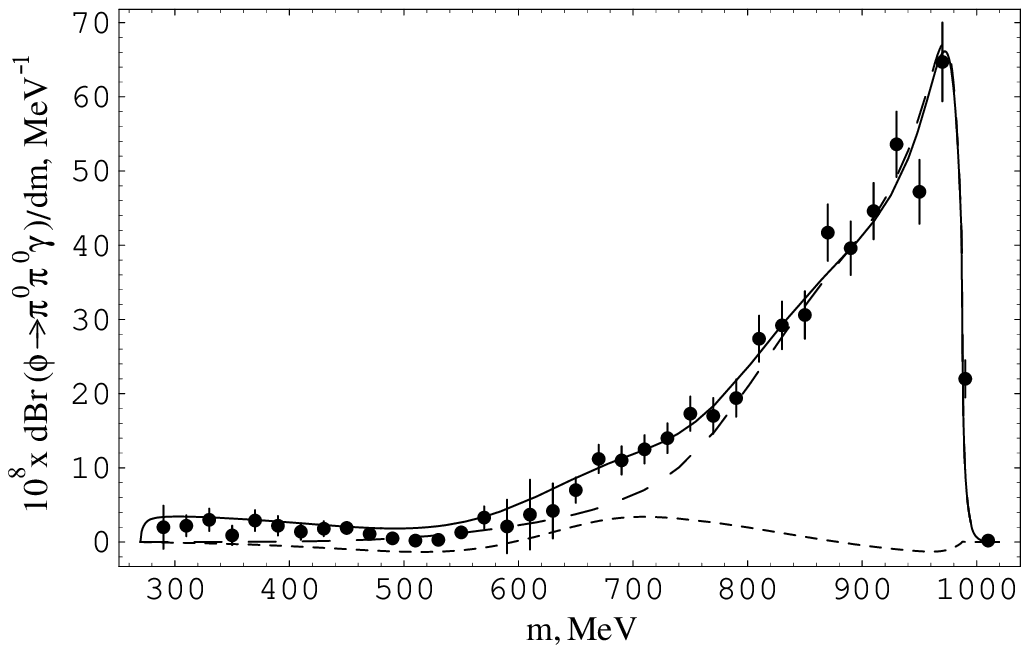}}
\end{tabular}\end{center}\vspace{-4mm}\caption{\footnotesize{The
left (right) plot shows the fit to the KLOE data for the $\pi^0\eta$
($\pi^0 \pi^0$) mass spectrum in the $\phi\to\gamma\pi^0\eta$
($\phi\to\gamma\pi^0\pi^0$) decay caused by the $a_0(980)$
($\sigma(600)+f_0(980)$) production through the $K^+K^-$ loop
mechanism.}}\end{figure}

\vspace{2mm}

\noindent\textbf{7\ \ Scalar Nature and Production Mechanisms in
\boldmath{$\gamma\gamma$} collisions \cite{A8209}}

Twenty seven years ago we predicted the suppression of
$a_0(980)\to\gamma\gamma$ and $f_0(980)\to\gamma\gamma$ in the
$q^2\bar q^2$ MIT model,
$\Gamma_{a_0\to\gamma\gamma}\sim\Gamma_{f_0\to\gamma\gamma}\sim
0.27\,\mbox{keV}$. Experiment supported this prediction.

Recently the experimental investigations have made great qualitative
advance. The Belle Collaboration  published data on
$\gamma\gamma\to\pi^+\pi^-$, $\gamma\gamma\to\pi^0\pi^0$, and
$\gamma\gamma\to\pi^0\eta$, whose statistics are huge
\cite{Mo07Ue0809}, see Fig. 5. They not only proved the theoretical
expectations based on the four-quark nature of the light scalar
mesons,  but also have allowed to elucidate the principal mechanisms
of these processes. Specifically, the direct coupling constants of
the $\sigma(600)$, $f_0(980)$, and $a_0(980)$ resonances with the
system are small with the result that their decays into
$\gamma\gamma$ are the four-quark transitions caused by the
rescatterings $\sigma(600)$\,$\to $\,$\pi^+\pi^-$\,$\to
$\,$\gamma\gamma$, $f_0(980)$\,$\to $\,$K^+K^-$\,$\to
$\,$\gamma\gamma$ and $a_0(980)$\,$\to $\,$K^+K^-$\,$\to
$\,$\gamma\gamma$ in contrast to the $\gamma\gamma$ decays of the
classic $P$ wave tensor $q\bar q$ mesons $a_2(1320)$, $f_2(1270)$
and $f'_2(1525)$, which are caused by the direct two-quark
transitions $q\bar q$\,$\to $\,$\gamma\gamma$ in the main. As a
result the practically model-independent prediction of the $q\bar q$
model $g^2_{f_2\gamma\gamma}:g^2_{a_2\gamma\gamma}=25:9$ agrees with
experiment rather well. The two-photon light scalar widths averaged
over resonance mass distributions $\langle\Gamma_{f_0\to\gamma
\gamma}\rangle_{\pi\pi}$\,$\approx$\,0.19 keV,
$\langle\Gamma_{a_0\to\gamma \gamma}\rangle_{\pi\eta}$\,$\approx
$\,0.3 keV and $\langle\Gamma_{\sigma
\to\gamma\gamma}\rangle_{\pi\pi}$\,$\approx$\,0.45 keV. As to the
ideal $q\bar q$ model prediction
$g^2_{f_0\gamma\gamma}:g^2_{a_0\gamma\gamma}=25:9$, it is excluded
by experiment.

\begin{figure}\begin{center}\begin{tabular}{ccc}{\epsfysize=1.8in\epsfbox{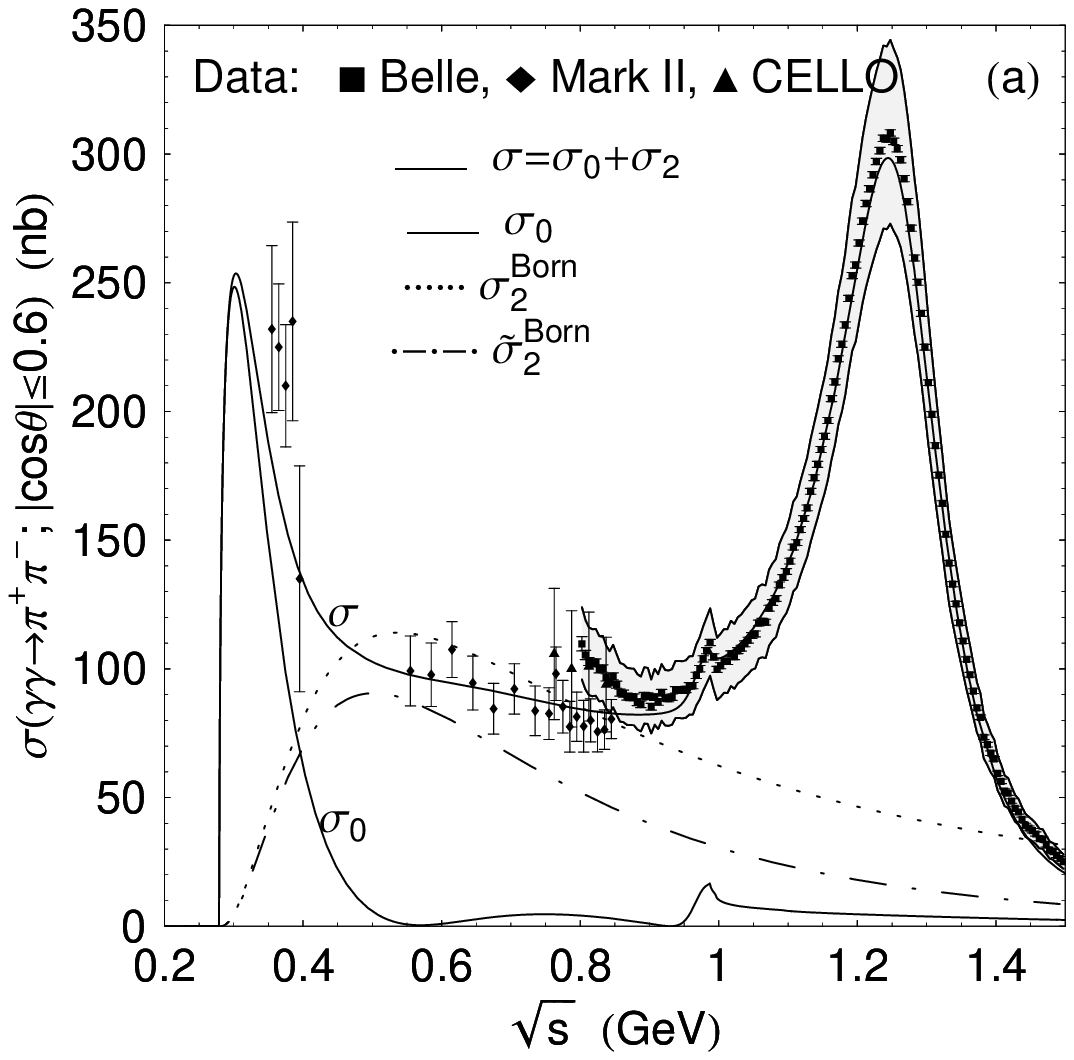}}\
\ & 
{\epsfysize=1.8in\epsfbox{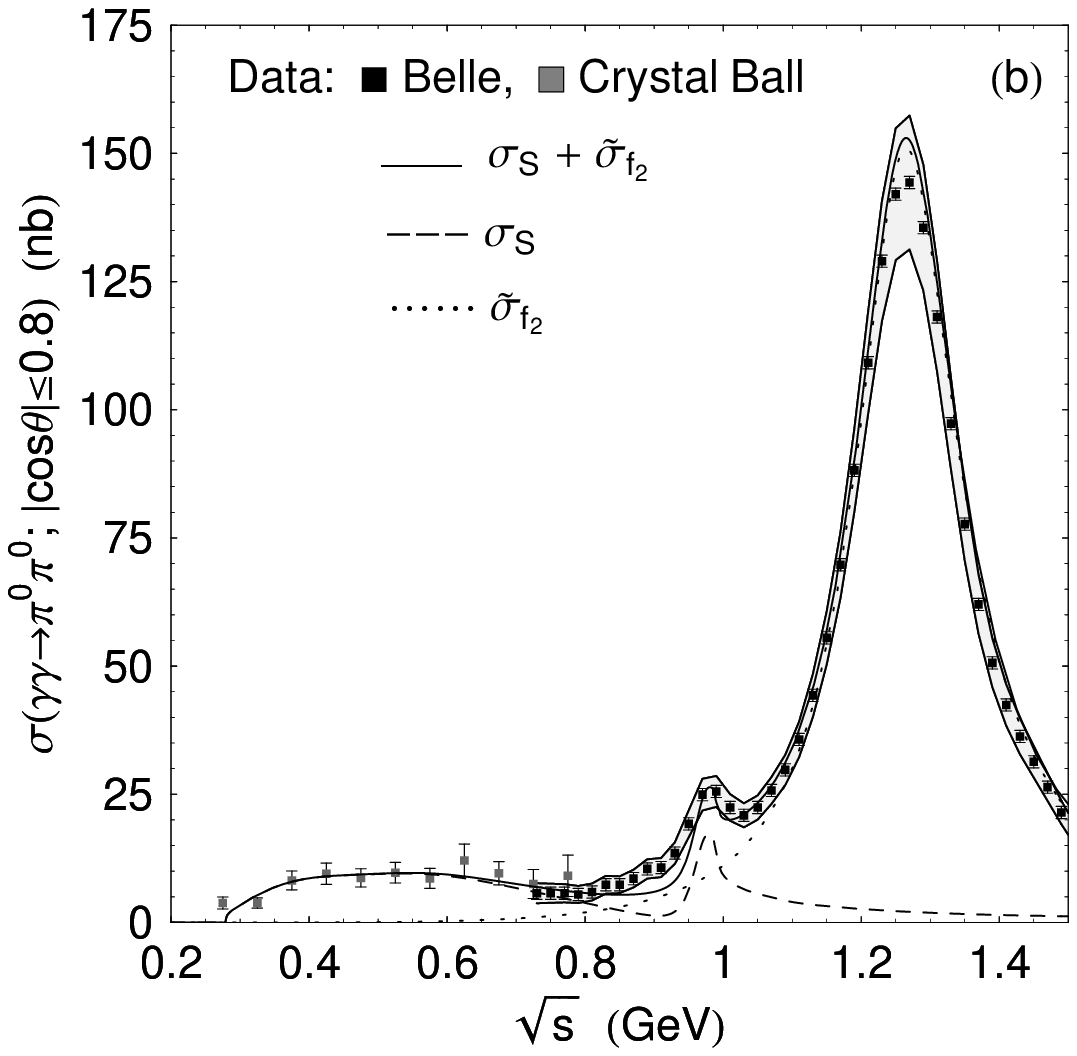}}\ \ &
{\epsfysize=1.8in\epsfbox{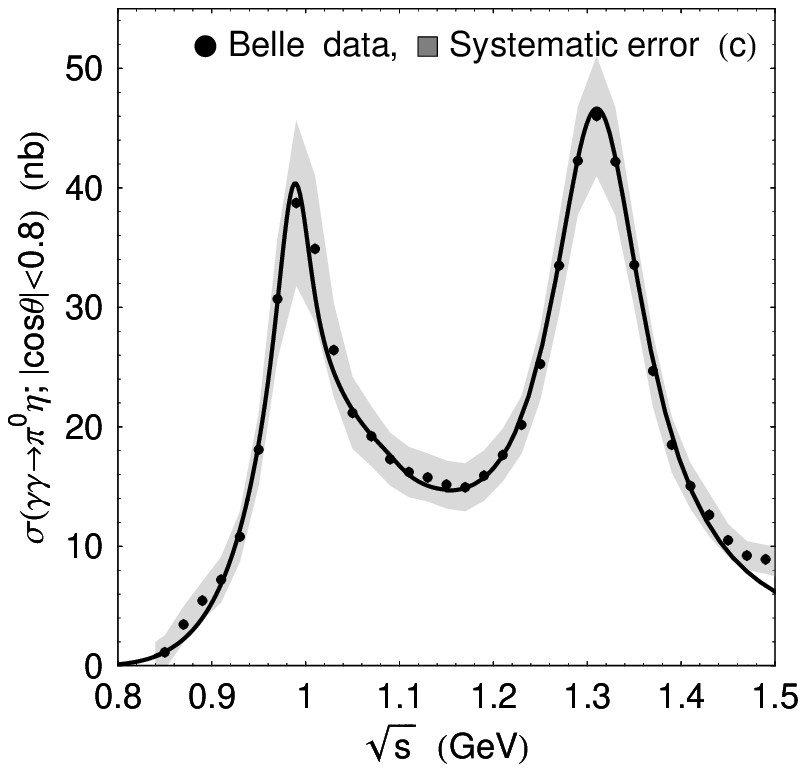}}\end{tabular}\end{center}\vspace{-4mm}\caption{{
\footnotesize Descriptions of the Belle data on
$\gamma\gamma$\,$\to$\,$\pi^+\pi^-$ (a),
$\gamma\gamma$\,$\to$\,$\pi^0\pi^0$ (b), and
$\gamma\gamma$\,$\to$\,$\pi^0\eta$ (c).}}\end{figure}

\vspace{2mm}

\noindent\textbf{8\ \ Summary of the Above
\cite{AS9407,A8907,A8209}}

\textbf{(i)} The mass spectrum of the light scalars, $\sigma (600)$,
$\kappa (800)$, $f_0(980)$, $a_0(980)$, gives an idea of their
$q^2\bar q^2$ structure. \textbf{(ii)} Both intensity and mechanism
of the $a_0(980)/f_0(980)$ production in the $\phi(1020)$ radiative
decays, the $q^2\bar q^2$ transitions $\phi\to K^+K^-\to\gamma
[a_0(980)$ $/f_0(980)]$, indicate their $q^2\bar q^2$ nature.
\textbf{(iii)} Both intensity and mechanism of the scalar meson
decays into $\gamma \gamma$, the $q^2\bar q^2$ transitions
$\sigma(600)\to\pi^+\pi^-\to\gamma\gamma$ and [$f_0(980) /a_0(980)
$]\,$\to K^+K^-\to\gamma\gamma$, indicate their $q^2\bar q^2$ nature
also.

\vspace{2mm}

\noindent\textbf{9\ \ \boldmath{The $a_0^0(980)-f_0(980)$} Mixing in
Polarization Phenomena \cite{ADS7904}}

The $a_0^0(980)-f_0(980)$ mixing as a threshold phenomenon was
discovered theoretically in 1979 in our work \cite{ADS7904}. Now it
is timely to study this phenomenon experimentally.\footnote{In Ref.
\cite{WZZ07} the search program of the $a_0^0(980)-f_0(980)$ mixing
at the $C/\tau$ factory has been proposed. Recently the VES
Collaboration published the data on the first effect of the
$a_0^0(980)-f_0(980)$ mixing, $f_1(1420)\to\pi^0 a_0^0(980)\to\pi^0
f_0(980)\to 3\pi$ \cite{Do08}, in agreement with our calculation
1981 \cite{ADS7904}.}

The main contribution originates from the
$a^0_0(980)$\,$\to$\,$(K^+K^-$\,$+$\,$K^0\bar
K^0)$\,$\to$\,$f_0(980)$ transition, see Fig. 6. Between the $K\bar
K$ thresholds $$|\Pi_{a_0f_0}(m)|\approx
\frac{|g_{a_0K^+K^-}g_{f_0K^+K^-}|}{16\pi}
\sqrt{\frac{2(m_{K^0}-m_{K^+})}{m_{K^0}}}\approx0.127
\frac{|g_{a_0K^+K^-}g_{f_0K^+K^-}|}{16\pi}\simeq
0.03\,\mbox{GeV}^2.$$ It dominates for two reason. i) It has the
$\sqrt{m_d-m_u}$\,$\sim$\,$\sqrt{\alpha}$ order. ii) The strong
coupling of the $a_0^0(980)$ and $f_0(980)$ to the $K\bar K$
channels, $|g_{a_0K^+K^-}g_{f_0K^+K^-}|/4\pi\simeq 1\,\mbox{GeV}^2$.

We noted in 2004 \cite{ADS7904} that the phase jump, see Fig. 6(b),
suggest the idea to study the $a_0^0(980)-f_0(980)$ mixing in
polarization phenomena. If a process amplitude with a suitable spin
configuration is dominated by the $a_0^0(980)-f_0(980)$ mixing then
a spin asymmetry of a cross section jumps near the $K\bar K$
thresholds. An example is the reaction
$\pi^-p_{\uparrow}\to\left(a_0^0(980)+f_0(980)\right)n\to
a_0^0(980)n\to\eta\pi^0n$, see Fig. 7. Performing the polarized
target experiments on the reaction $\pi^- p\to\eta\pi^0n$ at high
energy could unambiguously and very easily establish the existence
of the $a_0(980)-f_0(980)$ mixing phenomenon through the presence of
a strong ($\sim1$) jump in the normalized azimuthal spin asymmetry
of the  $S$ wave $\eta\pi^0$ production cross section near the
$K\bar K$ thresholds. In turn it could give an exclusive information
on the $a_0(980)$ and $f_0(980)$ coupling constants with the $K\bar
K$ channels, $|g_{a_0K^+K^-}g_{f_0K^+K^-}|/4\pi$.

\begin{figure} \centerline{\epsfxsize=3.5in
\epsfbox{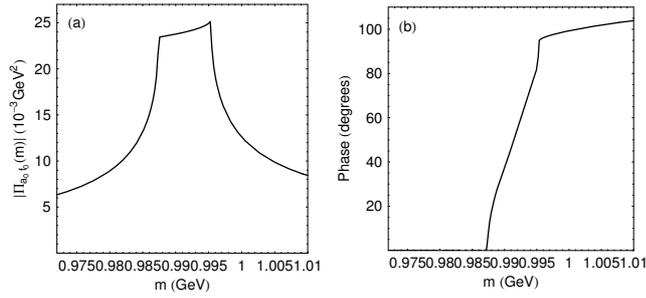}}\vspace{-2mm} \caption {{\footnotesize The
``resonancelike'' behavior of the modulus (a) and phase (b) of the
$a_0^0(980)-f_0(980)$ mixing amplitude
$\Pi_{a_0f_0}(m)$.}}\end{figure}

\begin{figure} \centerline{\epsfxsize=2in
\epsfbox{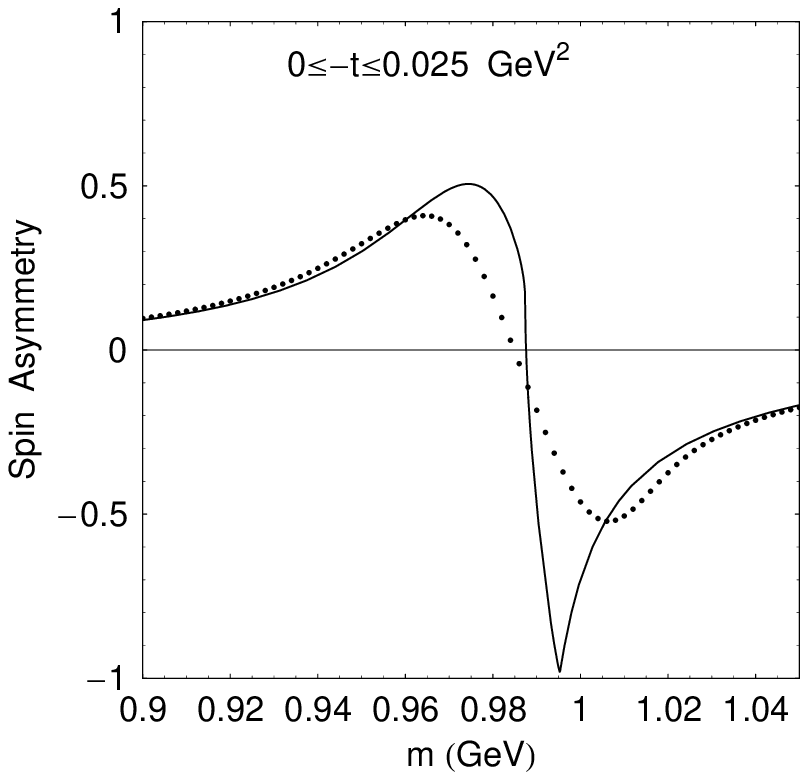}}\vspace{-2mm} \caption {{\footnotesize The
spin asymmetry in $\pi^-p_{\uparrow}$\,$\to$\,$\left (a_0^0(980)
+f_0(980) \right)n$\,$\to$\, $\eta\pi^0n$.}}\end{figure}

\vspace{4mm}  This work was supported in part by the RFFI Grant No.
07-02-00093 and by the Presidential Grant No. NSh-1027.2008.2 for
Leading Scientific Schools.

\footnotesize {}


\begin{thebibliography}{9}
\bibitem{AS9407} N.N.~Achasov and G.N.~Shestakov, Phys. Rev. {\bf D49} (1994)
                 5779; Phys. Rev. Lett. {\bf 99} (2007) 072001.
\bibitem{Ja77}   R.L.~Jaffe, Phys. Rev. {\bf D15} (1977) 267, 281.
\bibitem{A8907}  N.N.~Achasov and V.N.~Ivanchenko, Nucl. Phys. {\bf B315} (1989)
                 465; N.N.~Achasov and V.V.~Gubin, Phys. Rev. {\bf D56} (1997)
                 4084; Phys. Rev. {\bf D63} (2001) 094007; N.N.~Achasov, Nucl.
                 Phys. {\bf A728} (2003) 425; N.N.~Achasov and A.V.~Kiselev, Phys.
                 Rev. {\bf D68} (2003) 014006; Phys. Rev. {\bf D73} (2006) 054029.
\bibitem{A8209}  N.N.~Achasov, S.A.~Devyanin, and G.N.~Shestakov, Phys. Lett. {\bf 108B}
                 (1982) 134; Z.~Phys. {\bf C16} (1982) 55;
                 N.N.~Achasov and G.N.~Shestakov, Phys. Rev. {\bf D72} (2005)
                 013006; Phys. Rev. {\bf D77} (2008) 074020; Pisma Zh. Eksp. Teor.
                 Fiz. {\bf 88} (2008) 345; Pisma Zh. Eksp. Teor. Fiz. {\bf 90} (2009) 355.
\bibitem{Mo07Ue0809} T. Mori  et al., Phys. Rev. {\bf D75} (2007) 051101(R); J. Phys. Soc.
                 Jap. {\bf 76} (2007) 074102; S. Uehara et al., Phys. Rev. {\bf D78}
                 (2008) 052004; Phys. Rev. {\bf D80} (2009) 032001.
\bibitem{ADS7904}N.N.~Achasov , S.A.~Devyanin, and G.N.~Shestakov, Phys. Lett. {\bf B88} (1979)
                 367; Yad. Fiz. {\bf 33} (1981) 1337; N.N.~Achasov and G.N.~Shestakov, Phys. Rev.
                 Lett. {\bf 92} (2004) 182001; Phys. Rev. {\bf D70} (2004) 074015.
\bibitem{WZZ07}  J.-J.~Wu, Q.~Zhao, and B.S.~Zou, Phys. Rev. {\bf D75} (2007) 114012.
\bibitem{Do08}   V.~Dorofeev et al., Eur. Phys. J. {\bf A38} (2008) 149;
                 arXiv: 0712.2512 [hep-ex].
\end{thebibliography}
\end{document}